# Analytical 1-D dual-porosity equivalent solutions to 3-D discrete single-continuum models. Application to karstic spring hydrograph modelling.

F. Cornaton, P. Perrochet

*Centre of Hydrogeology, University of Neuchâtel. Emile-Argand 11, 2007 Neuchâtel, Switzerland*
*E-mail address:* fabien.cornaton@unine.ch

**Abstract**

One dimensional analytical porosity-weighted solutions of the dual-porosity model are derived, providing insights on how to relate exchange and storage coefficients to the volumetric density of the high-permeability medium. It is shown that porosity-weighted storage and exchange coefficients are needed when handling highly heterogeneous systems – such as karstic aquifers – using equivalent dual-porosity models. The sensitivity of these coefficients is illustrated by means of numerical experiments with theoretical karst systems. The presented 1-D dual-porosity analytical model is used to reproduce the hydraulic responses of reference 3-D karst aquifers, modelled by a discrete single-continuum approach. Under various stress conditions, simulation results show the relations between the dual-porosity model coefficients and the structural features of the discrete single-continuum model. The calibration of the equivalent 1-D analytical dual-porosity model on reference hydraulic responses confirms the dependence of the exchange coefficient with the karstic network density. The use of the analytical model could also point out some fundamental structural properties of the karstic network that rule the shape of the hydraulic responses, such as density and connectivity.

## 1. Introduction

Based on the definition of the first mathematical model developed by Barenblatt et al. (1960) for dual-porosity systems, many different schemes have been proposed in order to better describe the hydraulic behaviour of such reservoirs. The basic concept of the dual-porosity model is that the fractured rock consists of two overlapping continua in hydraulic interaction : a matrix continuum of low-permeability, primary porosity and a fracture continuum of high-permeability, secondary porosity. Hydrodynamic processes are thus controlled by a system of two partial differential equations. In literature, two main approaches can be distinguished: the first considers that flow between continua occurs under pseudo-steady state conditions (see e.g. Warren and Root, 1963), and the second considers a fully transient exchange between continua (see e.g. Kazemi, 1969). Fully transient exchange models are more sophisticated from the physical and mathematical point of view than the models based on steady-state transfer functions. However, Moench (1984) showed that pseudo steady-state block/fracture exchange models could be described as a particular case of unsteady transfer models by means of fracture skin parameters. In his analysis, Moench (1984) provided a theoretical justification for the use of equivalent pseudo steady-state flow models, where both overlapping continua have the same geometry. Since the main source of uncertainty in the prediction of the hydraulic behaviour of fractured aquifers lies in the distribution of high-permeability zones, those kinds of models do not require the definition of a block geometry, as in transient models. The flux exchanged between continua $q_{ex}$ is ruled by a lumped parameter, as defined by Barenblatt et al. (1960), according to the equation $q_{ex} = \pm\alpha(H_m - H_f)$, where $H_m$ and $H_f$ are the respective average hydraulic heads in the matrix continuum and the fracture continuum, and $\alpha$ is the exchange coefficient. Following the model of Warren and Root (1963), $\alpha$ relates to the fractured rock geometry and its hydraulic properties.

Karstic aquifers are commonly schematised by a mostly unknown high-permeability channel network which is embedded in a low-permeability fractured limestone volume, and is well connected to a discharge area. In such carbonate aquifers, which have often been assimilated to dual-porosity systems (Drogue, 1969; Teutsch, 1988; Mohrlok, 1996; Mohrlok and Sauter, 1997; Mohrlok and Teutsch, 1997), the geometry of the channel network governs the global hydraulic response but is never well known a priori. In a dual-porosity model of karstic aquifer, the primary porosity represents the matrix volumes (fissured limestone volumes of low-permeability) and the secondary porosity represents the karstic network (channel network of high-permeability). The two model components are therefore linked to this duality in the hydraulic characteristics.

The purpose of this paper is first to provide 1-D analytical solutions of the dual-porosity model with porosity-weighted continua, according to transient-type input boundary conditions. In a second step, we point out the relation between the exchange coefficient $\alpha$ and some geometrical and/or physical characteristics of the high-permeability zones in dual-porosity systems. To illustrate the application of the model, numerical experiments are carried out in order to simulate the hydraulic responses of several 3-D reference models which take into account the hydraulic characteristics of karstic aquifers. The calibration of the analytical model on reference data show the large effect of the karstic network structure on the hydraulic responses (spring hydrograph and hydraulic heads).

## 2. Analytical solutions for 1-D dual-porosity conduit flow

In this section, we propose 1-D solutions of the dual-porosity model (Fig. 1), with porosity-weighted continua. In this model, we assume zero matrix conductivity and first order exchange kinetics between porosities (pseudo-steady state exchange), ruled by the lumped $\alpha$ parameter. Solutions are given for a few typical transient input signals imposed as boundary condition.

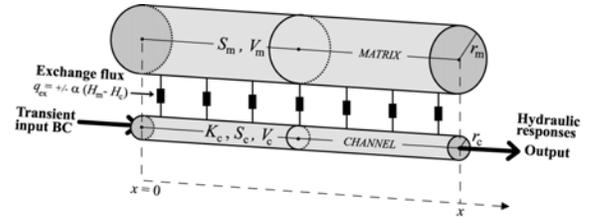

Fig. 1. Conceptual scheme of the analytical model.

### 2.1. Basic equations

The differential equation controlling flow in the channel continuum is described by the groundwater diffusion equation with source term :

$$K_c \frac{\partial^2 H_c}{\partial x^2} = S_c \frac{\partial H_c}{\partial t} + S_m \frac{\partial H_m}{\partial t} \qquad (1)$$

with $K_c$ and $S_c$ the hydraulic conductivity and storage coefficient of the high-permeability continuum (channel), and $S_m$ the storage coefficient of the low-permeability continuum (matrix). $H_m = H_m(x, t)$ and $H_c = H_c(x, t)$ are the hydraulic heads of the matrix continuum and the channel continuum respectively, $x$ being the distance.

In Eq. (1), the second term in the right hand side acts as source term accounting for contributions to and from the matrix continuum.

Let $\phi$ be the volumetric density of the channel continuum. $\phi$ is the ratio of the total channel volume to the apparent total volume of the system, and can also be assimilated to the channel total porosity. The matrix continuum volumetric density is thus $1-\phi$.

$$\phi = \frac{V_c}{V_a} \quad (2)$$

$$1 - \phi = \frac{V_m}{V_a} \quad (3)$$

where $V_c$ is the channel volume, $V_m$ is the matrix volume and $V_a$ is the apparent total volume.

Weighting the terms of Eq. (1) by $\phi$ and $1-\phi$ yields

$$\phi K_c \frac{\partial^2 H_c}{\partial x^2} = \phi S_c \frac{\partial H_c}{\partial t} + (1-\phi) S_m \frac{\partial H_m}{\partial t} \quad (4)$$

or

$$K_c \frac{\partial^2 H_c}{\partial x^2} = S_c \frac{\partial H_c}{\partial t} + \beta S_m \frac{\partial H_m}{\partial t} \quad (5)$$

where

$$\beta = \frac{(1-\phi)}{\phi} \quad (6)$$

When both continua are assimilated to tubes or pipes, Eq. (6) results in

$$\beta = \left(\frac{r_m}{r_c}\right)^2 \quad (7)$$

$r_c$ and $r_m$ being the channel and matrix radius respectively.

The expected solutions of Eq. (5) must satisfy the following boundary conditions:

$$H_c(0,t) = BC(t)$$

or $\quad -K_c \frac{\partial H_c}{\partial x}(0,t) = BC(t),$

and $\quad H_c(+\infty, t) = 0 \quad (8)$

with $BC(t)$ a transient input function at the boundary $x = 0$.

*2.2. Resolution*

Assuming first order exchanges between porosities, the matrix storage term in Eq. (5) can be expressed by

$$S_m \frac{\partial H_m}{\partial t} = -\alpha(H_m - H_c) \quad (9)$$

where $H_m = H_m(x, t)$, $H_c = H_c(x, t)$.

Applying a Laplace transform to Eq. (5) and Eq. (9) with the initial condition $H_c(x,0) = H_m(x,0) = 0$ yields

$$pS_c \langle H_c \rangle + p\beta S_m \langle H_m \rangle = K_c \frac{\partial^2 \langle H_c \rangle}{\partial x^2} \quad (10)$$

$$\langle H_m \rangle = \frac{\langle H_c \rangle}{1 + \frac{pS_m}{\alpha}} \quad (11)$$

The $p$-transformed $L(H_i)$ is denoted by $\langle H_i \rangle$.

By substitution of (11) in (10) the following partial differential equation is obtained:

$$\frac{\partial^2 \langle H_c \rangle}{\partial x^2} = A(p) \langle H_c \rangle \quad (12)$$

with

$$A(p) = \frac{p}{K_c} \left( S_c + \frac{\beta S_m}{1 + \frac{pS_m}{\alpha}} \right) \quad (13)$$

The solutions of Eq. (12) must satisfy the boundary conditions

$$\langle H_c \rangle(0, p) = \langle BC \rangle(p)$$

or $\quad -K_c \frac{\partial \langle H_c \rangle}{\partial x}(0, p) = \langle BC \rangle(p),$

and $\quad \langle H_c \rangle(+\infty, p) = 0 \quad (14)$

with $\langle BC \rangle(p)$ the $p$-transformed boundary condition $BC(t)$ at $x = 0$.

The general solution of Eq. (12) reads

$$\langle H_c \rangle(x,p) = \frac{1}{2}\Big( \big[\langle F_1\rangle(p)+\langle F_2\rangle(p)\big]e^{x\sqrt{A(p)}} \\ + \big[\langle F_2\rangle(p)-\langle F_1\rangle(p)\big]e^{-x\sqrt{A(p)}} \Big) \quad (15)$$

where $\langle F_1\rangle(p)$ and $\langle F_2\rangle(p)$ account for the boundary conditions.

To satisfy the condition $\langle H_c\rangle(+\infty,p)=0$, it is clear that $\langle F_1\rangle(p) = -\langle F_2\rangle(p)$. To satisfy the condition $\langle H_c\rangle(0,p) = \langle BC\rangle(p)$, then $\langle F_2\rangle(p) = \langle BC\rangle(p)$. Finally, Eq. (15) results in

$$\langle H_c\rangle(x,p) = \langle BC\rangle(p) e^{-x\sqrt{A(p)}} \quad (16)$$

Using a head Dirac boundary condition $\langle BC\rangle(p) = \langle \delta\rangle(p) = 1$, the solution of Eq. (12) is

$$\langle H_c\rangle(x,p) = e^{-x\sqrt{A(p)}} \quad (17)$$

which corresponds to the channel transfer function, and the transformed Darcy flux is

$$\langle q_c\rangle(x,p) = -K_c \frac{\partial \langle H_c\rangle}{\partial x} = K_c\sqrt{A(p)}\, e^{-x\sqrt{A(p)}} \quad (18)$$

Under the flux Dirac boundary condition $-K_c \left.\frac{\partial \langle H_c\rangle(x,p)}{\partial x}\right|_{x=0} = \langle\delta\rangle(p) = 1$, the transformed hydraulic head in the channel reads

$$\langle H_c\rangle(x,p) = \frac{e^{-x\sqrt{A(p)}}}{K_c\sqrt{A(p)}} \quad (19)$$

and the transformed flux is

$$\langle q_c\rangle(x,p) = e^{-x\sqrt{A(p)}} \quad (20)$$

which equals the transfer function of the system as defined by Eq. (17)

*2.3. Channel connections using the dual-porosity transfer function model*

Let $\langle TR\rangle(x,p)$ be the transformed transfer function of our dual-porosity system, explicitly defined by

$$\langle TR\rangle(x,p) = e^{-x\sqrt{\left(\frac{S_c}{K_c}+\frac{\beta\alpha S_m}{K_c(\alpha+pS_m)}\right)p}} \quad (21)$$

Under any Laplace $p$-transformed flux boundary condition $\langle BC\rangle(p) = -K_c\frac{\partial \langle H_c\rangle(0,p)}{\partial x}$ (e.g. see table 1), the solution of Eq. (12) reads

$$\langle H_c\rangle(x,p) = \langle BC\rangle(p)\frac{1}{K_c}\int_x^{+\infty}\langle TR\rangle(u,p)du \quad (22)$$

for the channel hydraulic head, and the transformed flux is

$$\langle q_c\rangle(x,p) = \langle BC\rangle(p)\langle TR\rangle(x,p) \quad (23)$$

The hydraulic head in the matrix continuum can thus be assessed by making use of Eq. (11)

$$\langle H_m\rangle(x,p) = \frac{\langle H_c\rangle(x,p)}{1+\frac{pS_m}{\alpha}} \quad (24)$$

and the flow rate is given by $\langle Q_c\rangle(x,p) = \langle q_c\rangle(x,p)\pi r_c^2$.

The flow model described by Eq. (22) and Eq. (23) can be regarded as a 'dual-porosity transfer function model', as the transformation of an arbitrary transient signal at the input boundary of the system is achieved by means of impulse response function (transfer function) convolution.

Table 1
Example of typical flux boundary conditions.

| BC type | Formula in the Laplace domain |
|---|---|
| (1) $n$ *step-flux functions* | $\sum_{j=1}^{n}\left[\dfrac{q_{oj}}{p}\left(e^{-p\sum_{i=1}^{j}(\tau_i+T_{i-1})} - e^{-p\sum_{i=1}^{j}(\tau_i+T_i)}\right)\right]$ |
| (2) $n$ *triangular flux functions* | $\sum_{i=1}^{n}\dfrac{q_{oi}e^{-pu_i}\left(t_{2i}-t_{2i-1}+t_{2i-1}e^{-pt_{2i}}-t_{2i}e^{-pt_{2i-1}}\right)}{p^2 t_{2i-1}(t_{2i}-t_{2i-1})}$ |

(1) $\tau_i$ is the time elapsed before the step-flux $i$ starts and $T_i$ is the injection duration of intensity $q_{oi}$.
(2) $u_i$ is the time elapsed before the triangular signal $i$ starts, $t_{2i-1}$ the time elapsed when the maximum flux $q_{0i}$ is reached, and $t_{2i}$ the total duration of the signal $i$.

Finally, for any Laplace $p$-transformed flux boundary condition, the convolution with the transfer function integral with respect to $x$ yields the hydraulic head solution, and the convolution with the transfer function itself yields the specific flux solution. This property allows the connection of several channels in terms of fluxes, as illustrated by Fig. 2.

At the connection of two channels, the corresponding flow rates can be summed and become the upstream signal for the next channel section. The final signal is thus simply obtained by successive convolution of the transfer function of a specific channel section by its upstream boundary condition.

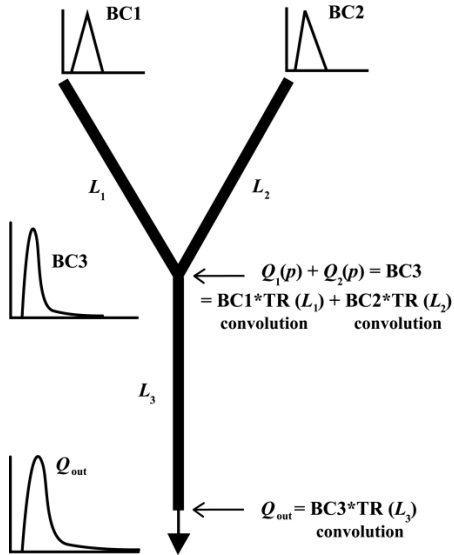

Fig. 2. Channel connection and resulting signal.

The Laplace transform technique has been widely used to assess analytical solutions of the dual-porosity model (see e.g. Warren and Root, 1963; Kazemi, 1969; Moench, 1984; Mohrlok, 1996). Mohrlok (1996) provided a 1-D solution in the Laplace domain and was able to calculate the inverse transformation of the Laplace transform without the requirement of a numerical inversion. However, the model proposed by this author differs from the one presented in this paper as no transient boundary condition at $x = 0$ is used. Moreover, if the time solution for a single channel system may be assessed, this would not be the case for the solutions of systems with multiple connected channels, solutions which can easily be derived by staying in the Laplace domain by means of convolutions.

## 3. Parameter discussion

### 3.1. Flux exchange coefficient

The dimension of the 3-D exchange coefficient $\alpha$ is [1/L/T]. For a 1-D model, the dimension is [L/T]. Considering the incremental volume $\pi r_c^2 dx$ in the channel, the exchange flux $q_{ex}$ [L/T] between porosities seeping through the surface $2\pi r_c dx$ can be formulated according to

$$q_{ex} = -\alpha [H_m(t) - H_c(t)] \frac{r_c}{2} \quad (25)$$

Moreover, using Darcy's law to evaluate $q_{ex}$ yields

$$q_{ex} = -K_m \frac{H_m(t) - H_c(t)}{\varepsilon r_m} \quad (26)$$

where $K_m$ is the matrix hydraulic conductivity [L/T] and $\varepsilon$ is a factor multiplying $r_m$ (the distance allowing to evaluate the gradient being unknown a priori).

From Eq. (25) and Eq. (26) one can write

$$\alpha = \frac{2}{\varepsilon} \frac{K_m}{r_c r_m} \quad (27)$$

If $\alpha_0$ is the 1-D exchange coefficient, then

$$\alpha_0 = \pi r_c^2 \alpha = \frac{2\pi}{\varepsilon} \frac{K_m}{\sqrt{\beta}} \quad (28)$$

Fig. 3 shows the behaviour of the simulated hydraulic head and flow rate according to the variations of the coefficient $\alpha$. The input is a symmetric triangular flux function (see Table 1). The more the value of $\alpha$ decreases, the steeper the peak amplitude of the global response, and also the steeper the decrease of the depletion curve. The output flux shows a depletion curve decreasing slower for increasing values of $\alpha$, involving a more and more important base-flow. This is due to the fact that when the transfer between porosities is high, during the stress event a big amount of water is stored in the matrix porosity. This volume is restored to the channel during a slow depletion.

For high magnitudes of $\alpha$ (e.g. with $\alpha = 1.10^{-8}$ on Fig. 3) the output signal is shifted, which is a consequence of the term under the square-root in Eq. (16), that tends to $p \frac{S_c + \beta S_m}{K_c}$ if $\alpha$ tends to infinity (high diffusion effect). Increasing values of $\alpha$ affect in the same way the hydraulic head response in the channel. On the contrary, when $\alpha$ increases the hydraulic head peak in the matrix continuum increases too. Inversions of hydraulic gradient, which are typical hydraulic processes occurring in karst systems, can well be observed during the rainfall event ($H_m < H_c$ during the stress period, $H_m > H_c$ during the recession period).

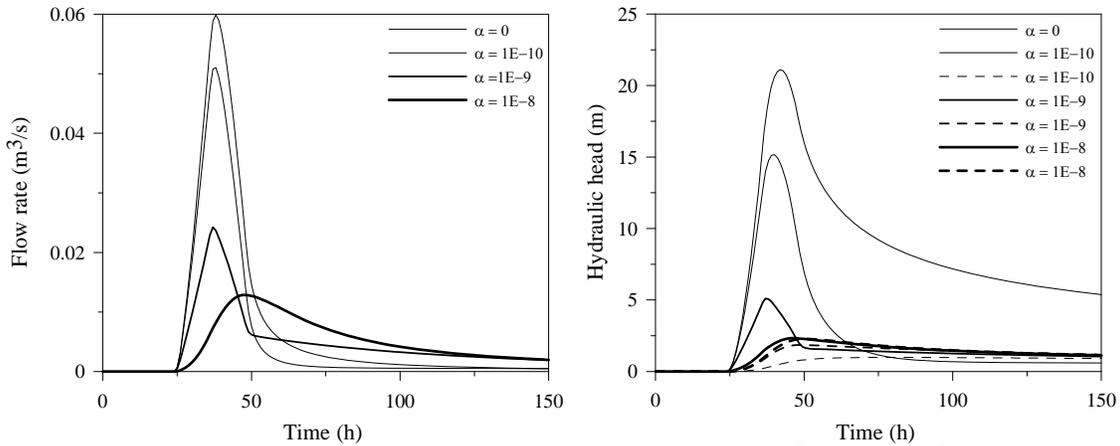

Fig. 3. Simulated output signals parameterised by coefficient $\alpha$ (m/s). $K_c = 10$ m$^3$/s, $S_c = 5.10^{-2}$ m, $S_m = 1.10^{-4}$ m, $\beta = 1.10^5$, $x = 1500$ m, $u = 24$ h, $t_1 = 12$ h, $t_2 = 24$ h, $q_0 = 0.1$ m$^3$/s. Bold lines : signals in the channel continuum; dashed lines : signals in the matrix continuum.

## 3.2. Porosity-weighting coefficient

In Barenblatt et al. (1960), the ratio between the fractured volume and the total apparent volume is neglected, compared to the matrix porosity. In this case, the total porosity of the dual-porosity system is assumed to be more or less equal to the matrix porosity. If $\beta = 1$ (i.e. $\phi = 0.5$), then Eq. (5) is equivalent to the common dual-porosity 1-D formulation without porosity-weighted continua (which is in fact similar to a weighting that equals ½), with 1st order exchange kinetics between porosities and assuming no flow in the matrix continuum. The porosity-weighted model thus represents a more general modelling approach, in which $0 < \phi < 1$, $\phi = 0.5$ being a particular case of the model. Note that if $\beta$ tends to infinity ($\phi \to 0$), then the hydraulic head tends to zero. If $\beta$ tends to zero ($\phi \to 1$), then the hydraulic head obeys the pure diffusion equation with the parameters of the channel continuum. Moreover, writing $\alpha' = \beta\alpha$ and $S'_m = \beta S_m$ yields

$$\langle TR \rangle (x, p) = e^{-x\sqrt{\left(\frac{S_c}{K_c} + \frac{\alpha' S'_m}{K_c(\alpha' + pS'_m)}\right)p}} \qquad (29)$$

Eq. (29) shows that including $\beta$ in the equations affects both the coefficients $\alpha$ and $S_m$.

Fig. 4 shows the behaviour of the simulated hydraulic head and flow rate according to the variations of the coefficient $\beta$. At a fixed exchange rate between porosities (constant $\alpha$), the coefficient $\beta$ will mainly rule the magnitude of the hydraulic responses (also involving modifications on the depletion part of the curves). For increasing values of this coefficient the spring hydrograph amplitude decreases. Both the hydraulic head amplitudes in the channel and in the matrix decrease too. This is the consequence of the fact that this coefficient affects in the same way the two coefficients $\alpha$ and $S_m$. At this point, one can see that introducing the coefficient $\beta$ in a dual-porosity model may be fruitful to calibrate both flow and head responses. In fact, Cornaton (1999) showed that using dual-continuum models with continua of equal volumes is inappropriate to calibrate both heads and spring hydrographs on data resulting from reference 3-D discrete single-continuum models of karstic aquifers.

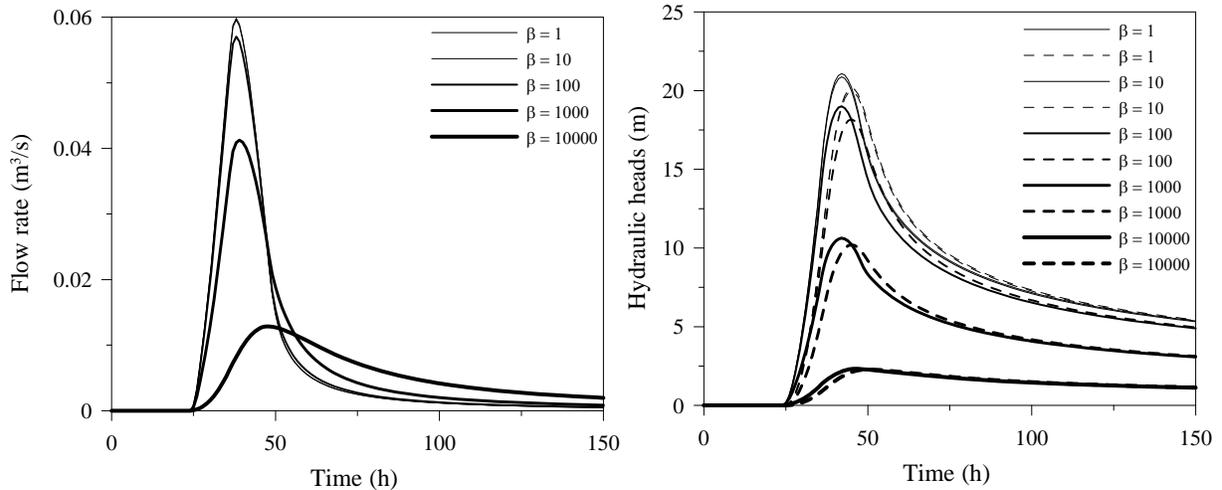

Fig. 4. Simulated output signals parameterised by coefficient $\beta$ (-). $K_c = 10$ m$^3$/s, $S_c = 5.10^{-2}$ m, $S_m = 1.10^{-4}$ m, $\alpha = 1.10^{-7}$ m/s, $x = 1500$ m, $u = 24$ h, $t_1 = 12$ h, $t_2 = 24$ h, $q_0 = 0.1$ m$^3$/s. Bold lines : signals in the channel continuum; dashed lines : signals in the matrix continuum. Note that the difference between the curves $\beta = 1$ and $\beta = 10$ is not sufficiently significant as to appear explicitly on the graph.

*3.3. Channel conductive parameter*

As flow in the model is laminar, the analogy between Poiseuille's and Darcy's laws suggests the following relation between the channel radius $r_c$ and the channel conductive parameter $K_c$ [$L^3/T$]

$$K_c = \frac{\pi \gamma}{8\mu} r_c^4 \qquad (30)$$

with $\gamma$ [M.L$^{-2}$.T$^{-2}$] and $\mu$ [M.L$^{-1}$.T$^{-1}$] the specific weight and dynamic viscosity of water.

*3.4. One dimensional storage coefficient*

Let $E_c$ [M.L$^{-1}$.T$^{-2}$] and $E_w$ [M.L$^{-1}$.T$^{-2}$] be the compressibility coefficients of a 1-D channel section and water respectively, $\theta$ [-] the channel porosity and $\gamma$ [M.L$^{-2}$.T$^{-2}$] the specific weight of water. The 1-D storage coefficient $S_c$ [L] is the specific storage coefficient [L$^{-1}$] multiplied by the channel cross-sectional area, that is

$$S_c = \gamma \left( \frac{\theta}{E_w} + \frac{1}{E_c} \right) \pi r_c^2 \qquad (31)$$

The $S_c$ [L] coefficient corresponds to the volume of liberated/stored water per unit of channel length, and per unit variation of hydraulic head. Neglecting the skull deformation and taking $\theta = 1$ for the channel porosity results in

$$S_c = \frac{\gamma}{E_w} \pi r_c^2 \qquad (32)$$

Enforcing Eq. (30) and Eq. (32) allows the simulation of typical karstic systems spring hydrographs. These hydrographs show a rapid and high intensity peak, followed by rapid decrease of the flow rate and a steep decrease of the depletion curve. However, straight application of Eq. (30) and Eq. (32) gives good results provided the two porosities have approximatively the same relative volumes ($\beta \approx 1$, $\phi \approx 0.5$). For larger values of $\beta$, Eq. (32) yields too high diffusivity values, transferring the input signal quasi-instantaneously. In this case, artificial values of channel storage coefficients are required, 100 to 1000 times higher, in order to match the experimental results.

This can be understood by considering that during a rainfall event, a big volume of water can be stored in the high-permeability network itself, by means of multiple permeability/porosity systems of connected open fractures, and other conduits. During the recession period, this amount must also participate in the volume of the low depletion part of the hydrograph, water being re-distributed within the higher permeability zones of the network, all the way to the spring. This process is illustrated in Fig. 5, showing the hydraulic steps during a rainfall event. During the recharge periods, the karstic network storage is accommodated by means of vertical conduits inside which the free water level rises. Water is re-collected by the main drains during the recession period. That is why the actual storage coefficient of the channels has to be artificially increased by orders of magnitude with respect to the value given by Eq. (32). This points out the fact that the depletion part of the hydrograph is not only an image of the low permeable volumes, but is also fully dependent on the channel network structure (density, connectivity, total volume), as already stated by Kiraly and Morel (1976). As a matter of fact, the pure exponential extremity of the depletion curve does not give any information on the hydraulic properties of the low permeable volumes, but rather on the resulting hydraulic properties of the set channel network/matrix volumes. Nevertheless, matrix specific yields and transmissivities are nowadays often still derived from the slope of the recession curve by fitting the data with an exponential function (Shevenell, 1996; Baedke and Krothe, 2001).

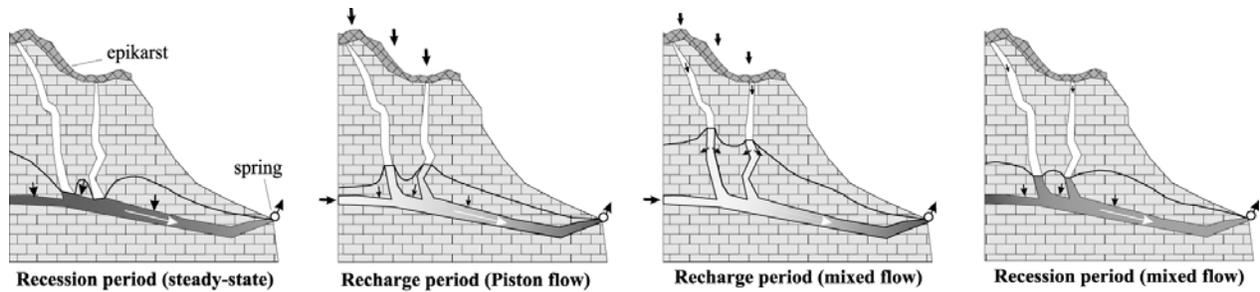
Fig. 5. Illustration of the storage process by the channel network during a rainfall event (modified after Grasso, 1998).

## 4. Sensitivity analysis

The organized heterogeneity of karst systems may be schematised by a mostly unknown high-permeability channel network which is embedded in a low-permeability fractured limestone volume, and is well connected to a discharge area, or karst springs. The duality of karst aquifers is a direct consequence of this structure (Kiraly, 1988, 1998) : duality of the infiltration processes (diffuse infiltration into the low-permeability volumes, concentrated infiltration into the channel network), duality of the groundwater flow field (low flow velocities in the fractured volumes, high flow velocities in the channel network) and duality of the discharge conditions (diffuse seepage from the low-permeability volumes, concentrated discharge from the channel network at the karst springs).

As the available data on karst channel networks are very limited, the combined discrete channel single-continuum approach cannot be widely used. However, it represents a powerful tool for checking the adequacy of the interpretation schemes based on a simpler representation of karst aquifers, where the channel network does not appear explicitly (e.g. global methods as black-box or grey-box models, simple continuum or double-continuum approach). Karst aquifers are 3-D systems and cannot be reduced to 2-D objects without loosing important information on the infiltration processes and the distribution of hydraulic heads. Numerical experiments with a 3-D finite element model using the combined discrete channel single-continuum approach, and simulating the infiltration and groundwater flow processes in a highly simplified theoretical karst aquifer, allowed to show the existence or non-existence of an epikarst zone enhancing concentrated infiltration (Kiraly et al., 1995).

The dual-porosity approach has the drawback that the model parameters can only be determined by model calibration i.e., the model parameters cannot be related directly to physical field measurements. In the following, we analyse the parameters used when a 1-D dual-porosity analytical model correctly describes the hydraulic responses of a well-known 3-D karst system.

*4.1. Discrete channel single-continuum karst systems*

Theoretical karst systems were elaborated using the computer codes FEN, enhanced versions of the original code FEM301 (Kiraly, 1985). FEN simulates the steady-state and transient three-dimensional saturated groundwater flow by the finite element method. The structure of the channel network is introduced in a 3-D domain, by means of 1-D elements, following Kiraly (1979, 1985, 1988). With this known structure, it becomes possible to analyse the relations between the reference structure of the aquifer and the parameters of the dual-porosity model which correctly describes the simulated hydraulic responses. Three different models were built, according to three different channel network densities with symmetrical properties (Fig 6). A set of horizontal and vertical channels are connected, forming vertical drainage panels connected to a main horizontal upper drain. A single spring is simulated by applying a constant head at a downstream node. Hydraulic parameters in the 3-D models are homogenously distributed. Model area and volume are 6.8 km$^2$ and 2.72 km$^3$ respectively. Fig. 6 shows the complexity of the simulated potential distribution in the aquifers. As a matter of fact, the knowledge of the groundwater table level (e.g. measured in boreholes) would not allow any interpretation of the flow processes.

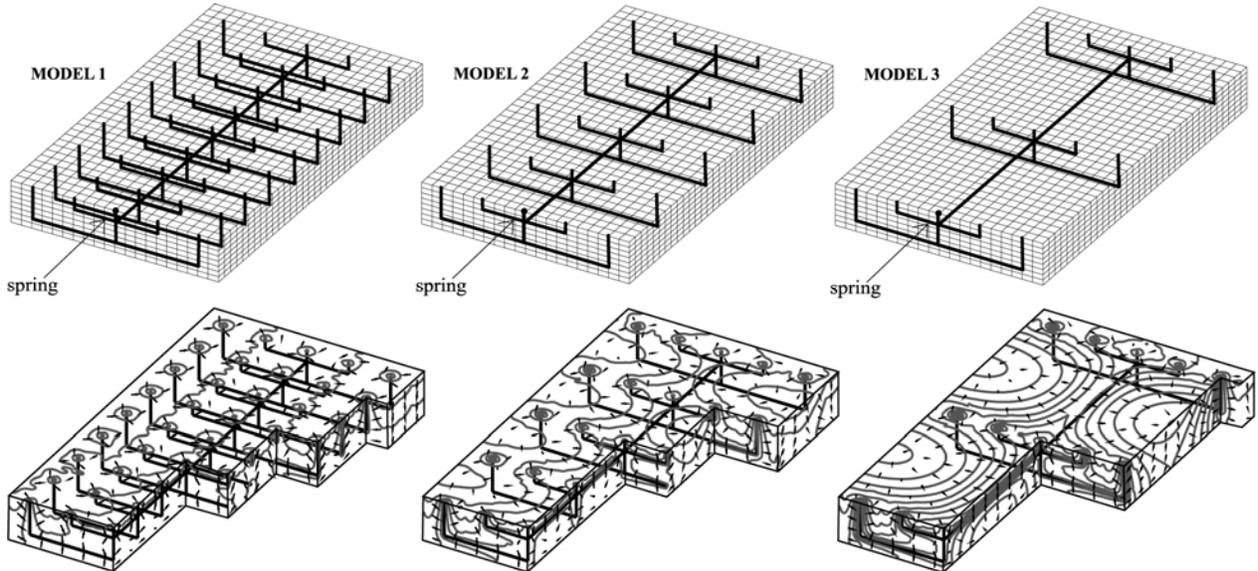

Fig. 6. (Up) Theoretical karst systems according to three channel network densities. Bold lines : karst channels embedded in a 3-D structure. (Below) Steady-state head solutions: equipotentials and flux vectors.

*4.2. Equivalent 1-D analytical dual-porosity model*

The calibration of a single channel dual-porosity model was rapidly abandoned as the results were too different from the reference data. Structurally equivalent 1-D analytical models were then constructed aiming at reproducing the right simulated reference internal responses and spring hydrographs. To perform this task, the structure of the karstic network of each discrete model was equivalently designed by using successive convolutions (as described on Fig. 2 and making use of Eq. (22) and Eq. (23) for the hydraulic head and flux, and of Eq. (2) in Table 1 for the boundary conditions). Fig. 7 illustrates the strategy which is used to create an equivalent dual-porosity model in the case of MODEL3. The signal produced by a panel $i$ is denoted by SIG$i$. This signal is routed through drain $j$, at the end of which it is noted BC$i$. For example at node 2, which connects the water routed by drain 1 and water produced by panel 2 (SIG2), the resulting signal is BC2 = TRD1 * SIG1 + SIG2, where TRD1 is the transfer function of drain 1, as defined by Eq. (21). This signal is then used as input for the routing convolution in drain 2, which, in turn, produces the signal reaching node 3. At node 4, the flow rate is equivalently calculated according to $Q_{out}$ = TR * BC3, where BC3 = TRD2 * BC2 + SIG3, and TR is the transfer function of the last channel section.

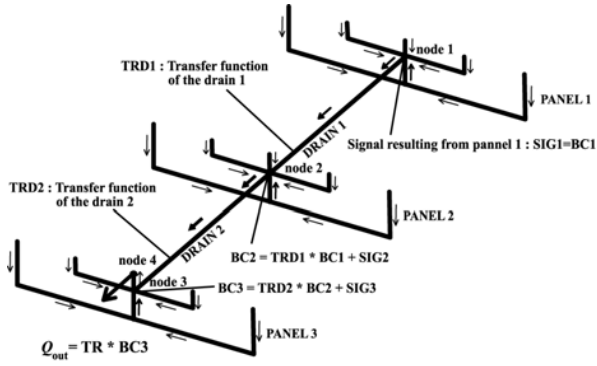

Fig. 7. Equivalent 1-D dual-porosity model in the case of MODEL3 using a cascading system of convolutions.

Finally, the structure of the karstic networks used in the discrete models is respected in terms of channel lengths and channel connectivity. Some parameters of the discrete models are taken as known values, namely the channel hydraulic conductivity and storage coefficient, and the matrix storage coefficient. The boundary conditions are respected (nodal flow rate at the extremity of each vertical channel, or sinkhole). In the discrete reference models, 100 % of the total infiltration volume are applied directly into the karstic network (100 % of concentrated infiltration).

In the equivalent 1-D dual-porosity model, drains and panels can have different parameters, as the drained volumes of matrix differ from one another: $\alpha$ and $\beta$ coefficients have then to be properly estimated and distributed.

To return into the time domain, the Laplace inversion was performed via the Crump's algorithm (Crump, 1976).

*4.3. Example of simulated hydrographs*

Several global responses of the reference karst aquifers were simulated, by using a sequence of one triangular symmetric function followed by one asymmetric infiltration function. The detail of a simulation example is given in table 2 and Fig. 8 illustrates the graphical result.

Table 2
Calibration example. Model parameters.

| 3-D numerical models | MODEL1 | MODEL2 | MODEL 3 |
|---|---|---|---|
| $K_c$ (m$^3$/s) | 10 | 10 | 10 |
| $S_c$ (m) | $5.10^{-3}$ | $5.10^{-3}$ | $5.10^{-3}$ |
| $K_m$ (m/s) | $1.10^{-6}$ | $1.10^{-6}$ | $1.10^{-6}$ |
| $S_m$ (1/m) | $5.10^{-5}$ | $5.10^{-5}$ | $5.10^{-5}$ |
| 1-D analytical models | MODEL 1 | MODEL 2 | MODEL 3 |
| *Panels* | | | |
| $\alpha$ (m/s) | $5.10^{-8}$ | $6.5.10^{-8}$ | $8.10^{-8}$ |
| $\beta$ (-) | $4.10^{3}$ | $2.25.10^{3}$ | $1.5.10^{3}$ |
| *Central drains* | | | |
| $\alpha$ (m/s) | $6.10^{-11}$ | $5.10^{-11}$ | $4.55.10^{-11}$ |
| $\beta$ (-) | $2.15.10^{5}$ | $2.5.10^{5}$ | $2.75.10^{5}$ |

To calibrate the three 1-D equivalent analytical models, the $\beta$ porosity weighting coefficient was first estimated by the evaluation of the drainable matrix volumes in the 3-D discrete models, during recession periods. This matrix volumes were then divided by the volume of the corresponding draining channels. As $K_c$, $S_c$ and $S_m$ were imposed in the analytical models, only the exchange coefficient $\alpha$ required a calibration. Fig. 8 shows the coherence between calibrated flow rates and hydraulic heads, in the case of MODEL3, where the temporal evolution of the head is 'measured' midway of drain 1 (see Fig. 7).

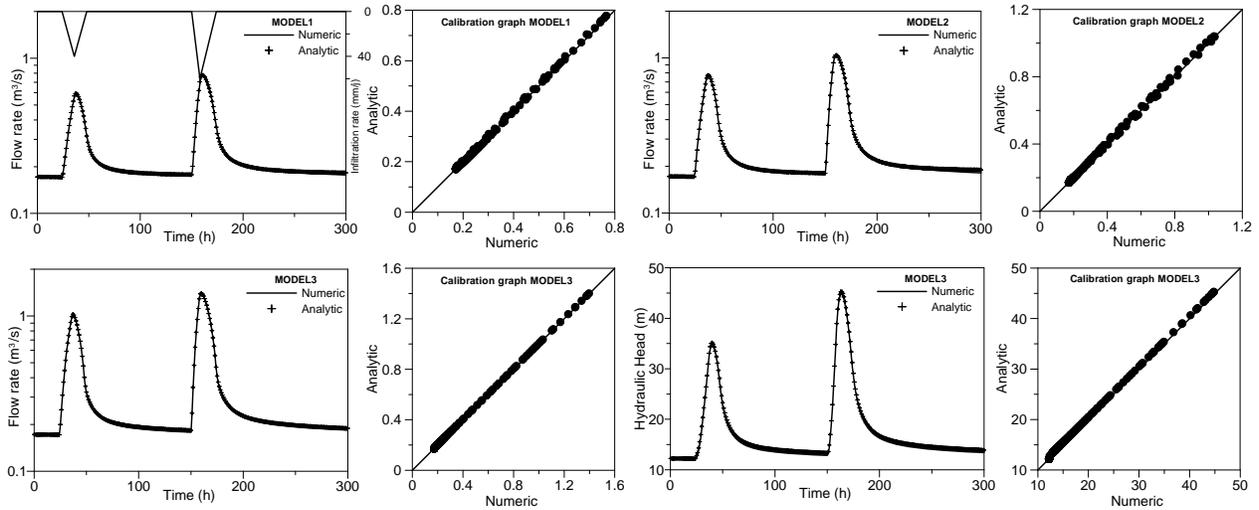
Fig. 8. Example of numerical/analytical calibrated fluxes, and calibrated hydraulic head for MODEL3.

The sensivity analysis confirmed the relation between the coefficients $\alpha$ and $\beta$, as defined by Eq. (28). However, the arbitrary constant $\varepsilon$ introduced in Eq. (28) could not be assessed through such an empirical analysis, mainly because of numerical restrictions on the simulated reference data, which highly depend on the level of refinement at the neighbourhood of the discrete 1-D elements. Note that discretisation around discrete elements essentially affects the depletion part of the reference hydraulic responses while the magnitudes of the simulated signals remain more or less the same for different levels of refinement.

## 5. Conclusion

It has been verified, by means of numerical experiments, that hydraulic responses of 3-D theoretical karstic systems can be obtained by 1-D porosity-weighted solutions of a dual-porosity system. Both the hydraulic heads in the channel network and spring hydrographs could be simulated by calibrating equivalent analytical dual-porosity models. Only the exchange coefficient $\alpha$ required a calibration process. To perform this task, the structure of the karstic network was the main characteristic of the reference aquifers that had to be taken into account. The combined analytical/numerical experiments showed that the network density and the number of channel connections within this network (channel network connectivity), are fundamental structural features that rule the shape and intensity of the spring hydrograph. The flux exchange coefficient in dual-porosity models is directly related to the karstic network density.

In order to simulate karstic shaped spring hydrographs with rapid responses to rainfall events, followed by low recession, both the 3-D numerical discrete single-continuum model and the 1-D analytical dual-porosity model require channel storage coefficient values 100 to 1000 times higher than the matrix storage coefficient. This fact can be explained by considering the potential storage capacity of the channel network in karst systems, accommodated by the high porosity zones (watering/dewatering of open fractures, vertical channels…) that can easily store a big amount of water during a rainfall event. This large additional storage process is neither dependent on the water compressibility nor the skull deformation.

Limitations of the presented dual-porosity model have to be clarified. Firstly, the use of this model to simulate karstic hydraulic responses is only valid for systems with fully concentrated infiltration (100 % of total seepage directly into the karstic network). Catchments of mature karstification, which are of great interest for water supply, represent an ideal situation for the application of the model. Secondly, as flow in the matrix continuum is neglected, the high-permeability network has to be the only water collector of the aquifer (i.e. concentrated discharge directly linked to the channel network). Finally, the position where the matrix hydraulic head is evaluated is unknown a priori, which makes the simulated matrix signals difficult to interpret.

Nevertheless, the use of porosity-weighted dual-porosity transfer function models allows to make inferences on the karstic network structure and its influence on the spring hydrograph. The presented analytical models, which remain valid for any transient boundary condition, combined with the possibility of channel connection, can be a practical tool for testing the karstic structure, the homogeneous/heterogeneous infiltration processes and parameters distribution effects on the hydraulic responses of highly heterogeneous systems such as karstic aquifers.

**Acknowledgements**

The authors would like to thank Professor Laszlo Kiraly for the use of his numerical code FEN, and for fruitful discussions on the subject of this paper.